\documentclass[aps,reprint,twocolumn,showpacs,preprintnumbers,amsmath,amssymb,nofootinbib,superscriptaddress,showkeys]{revtex4-1}

\usepackage{epsfig}

\begin{document}

\title{Perturbative Renormalizability of Chiral Two Pion Exchange \\
in Nucleon-Nucleon Scattering}
	\author{M. Pav\'on Valderrama}\email{m.pavon.valderrama@fz-juelich.de} 
\altaffiliation[present address: ]{Departamento de F\'{\i}sica Te\'orica and
Instituto de F\'{\i}sica Corpuscular (IFIC), Centro Mixto CSIC-Universidad de Valencia, Institutos de Investigaci\'on de Paterna, Aptd. 22085, E-46071 Valencia, Spain}
	\affiliation{Institut f\"ur Kernphysik and J\"ulich Center for Hadron Physics, Forschungszentrum J\"ulich, 52425 J\"ulich, Germany} 

\date{\today}

\begin{abstract} 
\rule{0ex}{3ex} 
We study the perturbative renormalizability of chiral two pion exchange
for the singlet and triplet channels within effective field theory,
provided that the one pion exchange piece 
of the interaction has been fully iterated.
We determine the number of counterterms/subtractions needed in order to
obtain finite results when the cut-off is removed,
resulting in three counterterms for the singlet channel and six
for the triplet.
The results show that perturbative chiral two pion exchange reproduce
the data up to a center-of-mass momentum of $k \sim 200-300\,\rm MeV$
in the singlet channel and $k \sim 300-400\,\rm MeV$ in the triplet.
\end{abstract}

\pacs{03.65.Nk,11.10.Gh,13.75.Cs,21.30.-x,21.45.Bc}
\keywords{Potential Scattering, Renormalization, Nuclear Forces, Two-Body System}

\maketitle

\section{Introduction}

The effective field theory formulation of nuclear forces~\cite{Beane:2000fx,Bedaque:2002mn,Epelbaum:2005pn,Epelbaum:2008ga} 
tries to exploit in a systematic manner 
the separation of scales between pion physics,
which is known to dominate at large distances, 
and short range physics in the two-nucleon system.
In Weinberg's original proposal~\cite{Weinberg:1990rz,Weinberg:1991um}
the chiral nucleon-nucleon potential is organized as a
power expansion (or counting) in terms of $Q$
\begin{eqnarray}
V_{\rm NN}(r) = V^{(0)}(r) + V^{(2)}(r) +
V^{(3)}(r) + \mathcal{O}(Q^4) \, , \label{eq:chiral-pot}
\end{eqnarray}
where $Q$ represents the low-energy scales of the system, usually
the momentum $p$ of the nucleons and the pion mass $m_{\pi}$.
The potential is then inserted into the Schr\"odinger or Lippmann-Schwinger
equation in order to obtain theoretical predictions
\cite{Ordonez:1993tn,Ordonez:1995rz,Kaplan:1996xu,Kaiser:1997mw,Kaiser:1998wa,Frederico:1999ps,Entem:2001cg,Entem:2002sf,Entem:2003ft,Epelbaum:2003gr,Epelbaum:2003xx,Epelbaum:2004fk,Nogga:2005hy,Valderrama:2005wv,PavonValderrama:2005uj,Krebs:2007rh,Higa:2007gz,Valderrama:2008kj,Valderrama:2010fb,Entem:2007jg,Borasoy:2006qn,Borasoy:2007vi,Yang:2007hb,Yang:2009kx,Yang:2009pn,Yang:2009fm}.
This prescription is usually referred to as Weinberg's counting.

The resulting chiral potentials turn out to be singular,
behaving at order $Q^{\nu}$ as $1/r^{3+\nu}$ in coordinate space for short
enough distances ($m_{\pi} r \ll 1$).
Therefore they need to be regularized in order to obtain well-defined results,
usually by introducing a cut-off in the computations plus the necessary
number of counterterms which ensure the renormalizability
of the scattering amplitude~\footnote{
It should be noted though that renormalization can be understood
in other ways, see the following paragraph.}.
This has been found to be in contradiction with Weinberg's power counting,
where the corresponding counterterms, determined by naive dimensional analysis,
are not able to render the theory renormalizable~\cite{Nogga:2005hy,Valderrama:2005wv,PavonValderrama:2005uj,Entem:2007jg}
(or generate chiral inconsistencies~\cite{Kaplan:1996xu},
prompting the KSW counting~\cite{Kaplan:1998tg,Kaplan:1998we}).
Consequently one is forced to make a decision: either to follow an
{\it a priori} power counting or require renormalizability.

The direct and practical choice is to follow Weinberg's original
counting unaltered,
leading to a framework amicable with large numerical computations,
which demystifies nuclear forces and enjoys an undisputed phenomenological
success~\cite{Entem:2003ft,Epelbaum:2004fk}.
The price to pay is that the cut-off must be fine tuned,
lying inside a narrow window, a situation which we regard
as unsatisfactory from a theoretical viewpoint.
Recently, based on the renormalization philosophy of
Lepage~\cite{Lepage:1997cs,Lepage:1989hf},
there has been interesting attempts to justify this particular
approach~\cite{Epelbaum:2006pt,Epelbaum:2009sd}.

On the contrary, if one strives for a more robust theoretical foundation,
one should be able to achieve cut-off independence.
The results from non-perturbative renormalization in the case of
singular interactions~\cite{Beane:2000wh,Nogga:2005hy,Valderrama:2005wv,PavonValderrama:2005uj,PavonValderrama:2007nu,Entem:2007jg,Entem:2009mf}
can be summarized as follows:
one counterterm is needed to renormalize a channel where the potential is
attractive and singular,
while channels where the potential is singular and repulsive
become insensitive to counterterms.
The first condition can lead to an alarming loss of predictive power,
as already at leading order (LO) there is an infinite number of
attractive singular channels.
The solution proposed in Ref.~\cite{Nogga:2005hy}
is to treat all partial waves with sufficiently
high angular momentum perturbatively at ${\rm LO}$,
a procedure which is supported by the analysis of Ref.~\cite{Birse:2005um}
~\footnote{An alternative solution has been recently 
proposed in Ref.~\cite{Valderrama:2010fb}.}.
The second condition is particularly problematic: in the triplet channel
the potential is attractive at ${\rm LO}$ but becomes repulsive at 
next-to-leading order (${\rm NLO}$),
resulting in an unbound deuteron 
at this order when the cut-off is removed~\cite{Valderrama:2005wv}.
As there is no way to predict what the sign of the interaction will be
at higher orders, this represents a continuous threat to the non-perturbative
renormalizability of the chiral potentials.
In addition, there exists the risk that non-perturbative renormalization
of the subleading pieces of the potential may lead to incompatibilities
with the chiral expansion~\cite{Epelbaum:2009sd}.
The previous issues can be avoided with the perturbative treatment of
the higher order pieces of the potential, which respects power counting
and renormalizability independently on whether the subleading
contributions are repulsive or attractive.
The problem is how to construct such a perturbation theory.

The purpose of this paper is to investigate the conditions under
which perturbative chiral two pion exchange (TPE)
can be renormalized
in order to extend the power counting proposal of
Nogga, Timmermans and van Kolck~\cite{Nogga:2005hy}
to subleading orders.
In the spirit of Refs.~\cite{Nogga:2005hy,Long:2007vp}, 
we use renormalizability as a guide to identify
the required short distance operators.
The technical meaning of renormalizability depends on whether we are
in a perturbative or non-perturbative context.
By perturbative renormalizability we refer to the elimination of all
negative  (positive) powers of the coordinate (momentum) space cut-off
in the observables.
In contrast, non-perturbative renormalizability deals with ambiguities
instead of divergences: the scattering amplitude of
an attractive singular interaction is finite but non-unique
and requires the inclusion of a counterterm for fixing the solution~\cite{Beane:2000wh,Valderrama:2005wv,PavonValderrama:2005uj}.
A particularly straightforward manner to fulfill the renormalization program
is to study the cut-off dependence of the amplitudes when the cut-off
is removed, as exemplified in Ref.~\cite{Nogga:2005hy}.
This should not be interpreted however as the necessity of eliminating the
cut-off in the computations: after the renormalization process,
the residual cut-off dependence of the amplitudes is in principle
a higher order effect, provided the cut-off lies
within a sensible range.

The perturbative techniques in this paper are directly based on 
those sketched in Ref.~\cite{Valderrama:2005wv},
and are equivalent to the momentum space perturbative
methods developed in Ref.~\cite{Long:2007vp}.
Here we use renormalized distorted wave Born Approximation (DWBA)
with the aim of constructing phase shifts.
Complementarily, the approach of Refs.~\cite{Birse:2003nz,Birse:2007sx,Birse:2010jr,Ipson:2010ah}
employs the DWBA techniques for ``deconstructing''
the phenomenological phase shifts, that is,
for extracting the corresponding short range physics
once the long range pion effects have been removed
and checking whether this short range interaction is consistent
with the specific power counting under consideration,
be it either Weinberg~\cite{Birse:2003nz}
or Nogga, Timmermans, van Kolck~\cite{Birse:2007sx,Birse:2010jr,Ipson:2010ah}.
Of particular interest is the recent deconstruction of
the $^1S_0$ singlet channel~\cite{Birse:2010jr}
which advances some of the results and conclusions of the present work.
The present approach differs however
from the finite cut-off perturbative set-up of
Ref.~\cite{Shukla:2008sp}, in which not all the operators
needed to obtain a renormalized results are included
as the previous work concentrates on analyzing the Weinberg counting.

The paper is organized as follows: in Section \ref{sec:singlet} we study
the perturbative renormalizability of the $^1S_0$ singlet channel
and determine the cut-off and momentum range for which
an acceptable description of the data is obtained.
We extend the previous results to the the $^3S_1-{}^3D_1$ triplet channel
in Section \ref{sec:triplet}.
The role of the cut-off within the present approach is analyzed 
in Section \ref{sec:cut-off}, and the relation with other approaches,
particularly the renormalization group analysis of Ref.~\cite{Birse:2005um},
is considered in detail in Section \ref{sec:approaches}.
Finally, we briefly summarize our results in \ref{sec:conclusions}.
The technical details of the perturbative treatment of chiral TPE
are explained in Appendices \ref{app:DWBA-singlet} and \ref{app:DWBA-triplet}.
Some of the $^1S_0$ singlet results from this paper have been advanced
in Ref.~\cite{Valderrama:2010aw}.

\section{Singlet Channel}
\label{sec:singlet}

The present perturbative treatment of chiral TPE
is based on distorted wave Born approximation.
For simplicity, we will only consider in detail the singlet case.
We can express the phase shifts as the following series 
\begin{eqnarray}
\delta(k; r_c) =
\delta^{(0)}(k; r_c) + \delta^{(2)}(k; r_c) + \delta^{(3)}(k; r_c) +
\mathcal{O}(Q^4) \, ,   \nonumber \\
\end{eqnarray}
which is ordered according to the counting of the finite-range piece of
the potential~\footnote{Strictly speaking, the leading order piece is
of order $Q^{-1}$ and not $Q^0$. However, for keeping the notation
simpler, we have just followed Eq.~(\ref{eq:chiral-pot}).}.
That is, power counting is now manifest in the amplitudes.
The ${\rm LO}$ phase shift $\delta^{(0)}$ is computed non-perturbatively
(and includes one counterterm~\footnote{Note that we are not considering here
chiral symmetry breaking terms separately.}),
while $\delta^{(2)}$ and $\delta^{(3)}$ are computed in first order
perturbation theory~\footnote{Second order perturbation theory is not needed
as the iteration of the ${\rm NLO}$ potential is of order $Q^5$.
}.
The corresponding expression for the perturbative phase shifts is
(see Appendix~\ref{app:DWBA-singlet})
\begin{eqnarray}
\frac{\delta^{(\nu)}(k; r_c)}{{\sin^2{\delta^{(0)}}}} &=& -
\frac{2\mu}{k}\,{\mathcal{A}^{(0)}(k; r_c)\,}^2\,
I^{(\nu)}_{{}^1S_0}(k; r_c) \, ,
\end{eqnarray}
where $\nu = 2, 3$
and the perturbative integral $I^{(\nu)}_{{}^1S_0}$ is defined as
\begin{eqnarray}
I^{(\nu)}_{{}^1S_0}(k; r_c) &=&
\int_{r_c}^{\infty}\,dr\,V^{(\nu)}(r)\,
{{u}_k^{(0)}\,}^2(r)\, .
\end{eqnarray}
In the previous formulae $\mu$ is the reduced mass,
$k$ the center of mass momentum,
$\mathcal{A}^{(0)}$ is a normalization factor,
which is taken to be unity at $k = 0$,
and $u_k^{(0)}$ is the ${\rm LO}$ reduced wave function 
in an energy independent normalization at the origin
(or at the cut-off radius $r_c$ if we are using a finite cut-off).
The asymptotic normalization of $u_k^{(0)}$ is determined by
$\mathcal{A}^{(0)}(k)\,u^{(0)}(k)
\to \sin{(k r + \delta^{(0)})} / \sin{\delta^{(0)}}$
for $r \to \infty$.

As can be easily checked, the perturbative integral diverges as
$1/r_c^{\nu + 2}$ as a consequence of the short distance behaviour of
the reduced wave function $u_k^{(0)}(r) \sim 1$ and 
the potential $V^{(\nu)}(r) \sim 1/r^{\nu + 3}$.
The divergences can be cured by making the adequate subtractions.
Due to the energy-independent normalization of $u^{(0)}_k$
at the origin, the terms in the $k^2$ expansion of
${u}_k^{(0)} = \sum_{n} {u}_{2n}^{(0)} k^{2n}$
are progressively less singular,
with ${u}_{2n}^{(0)} \sim r^{2 n}$ for $r \to 0$.
Expanding the previous integrals in terms of $k^2$ for $\nu = 2,3$
we have
\begin{eqnarray}
I^{(\nu)}_{{}^1S_0}(k; r_c) &=& I^{(\nu)}_0(r_c) + k^2\,I^{(\nu)}_2 (r_c) + 
k^4\,{I^{(\nu)}_4}(r_c) \nonumber \\ &+&  I^{(\nu)}_{{}^1S_0,R} (k; r_c) \, ,
\end{eqnarray}
where $I^{(\nu)}_{0, 2, 4}$ are the divergent pieces of the integral
and $I^{(\nu)}_{{}^1S_0,R}$ is the regular piece, 
as can be trivially checked.
Therefore three subtractions or counterterms are needed in order to renormalize
the perturbative results in the singlet.
The specific method employed is not important.
Here we modify the perturbative integral by adding three free parameters
which are to be fitted to the data
\begin{eqnarray}
\hat{I}^{(\nu)}_{{}^1S_0}(k; r_c) = \lambda^{(\nu)}_0 + 
\lambda^{(\nu)}_2 k^2 + \lambda^{(\nu)}_4 k^4 +
I^{(\nu)}_{{}^1S_0}(k; r_c) \, .
\end{eqnarray}
By assuming the short range physics to be parametrized
by an energy dependent delta-shell potential of the type
\begin{eqnarray}
V^{(\nu)}_C(r; r_c) &=& \frac{\mu\,}{2 \pi\, r_c^2}\,
\sum_{n} C^{(\nu)}_{2n}(r_c) k^{2 n} \, \delta(r-r_c) \, ,
\end{eqnarray}
we can easily relate the $\lambda^{(\nu)}_{2n}$ parameters to
the $C^{(\nu)}_{2n}$ counterterms by
\begin{eqnarray}
\lambda^{(\nu)}_{2n} = \frac{\mu\,}{2 \pi\,r_c^2}\,C^{(\nu)}_{2n}(r_c)
\,{u_0^{(0)}\,}^2(r_c) \, .
\end{eqnarray}
Equivalently, if one chooses to work in the momentum space formulation of
Ref.~\cite{Long:2007vp}, one could include the contact potential
$\langle p | V_C^{(\nu)} | p'\rangle =
C_0^{(\nu)} + C_2^{(\nu)} (p^2 + {p'}^2) + C_4^{(\nu)} (p^4 + {p'}^4)$.
In either case, the first free parameter, $\lambda^{(\nu)}_0$ ($C_0^{(\nu)}$),
is only used to absorb the $k = 0$ divergence of the perturbative integral
while its finite piece is redundant, as it only affects the zero energy
behaviour of the phase shifts which has already been fixed at ${\rm LO}$,
meaning that we need to fix two additional observables, for example
the effective range $r_0$ and the shape parameter $v_2$,
in order to determine the ${\rm NLO}$/${\rm N^2LO}$ results.
The number of counterterms agrees
with the corresponding one predicted in the RGA of Ref.~\cite{Birse:2005um},
where the power counting resulting from treating one pion exchange (OPE)
non-perturbatively was analyzed in detail,
and with the related deconstruction of Ref.~\cite{Birse:2010jr},
in which the short range physics for the singlet channel is determined
by removing the non-perturbative OPE and perturbative TPE effects
from the phenomenological phase shifts.
Note that in Ref.~\cite{Valderrama:2005wv} an incorrect number of counterterms
was determined due to an improper normalization.

\begin{figure}[htb]
\begin{center}
\epsfig{figure=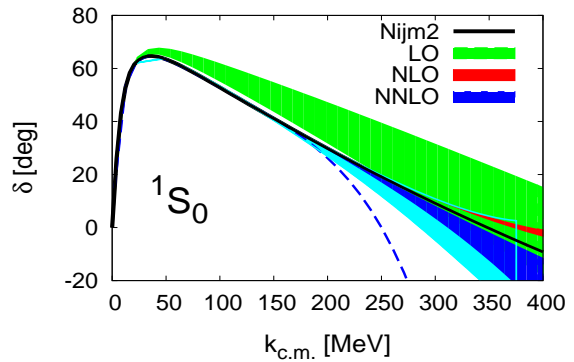,
	height=5.25cm, width=8.0cm}
\end{center}
\caption{(Color online)
Phase shifts for the $^1S_0$ channel with non-perturbative OPE and
perturbative TPE.
The non-perturbative OPE computation contains one counterterm which
is determined by fixing the $^1S_0$ scattering length,
$a_{0,s} = -23.74\,{\rm fm}$,
while the perturbative TPE computation contains 
a correction to the ${\rm LO}$ counterterm plus 
two additional counterterms which are used to fit 
the Nijmegen II phase shifts~\cite{Stoks:1994wp}
(equivalent to the Nijmegen PWA~\cite{Stoks:1993tb})
in the range $k = 40-160\,{\rm MeV}$.
The error bands are generated varying the cut-off
within the $0.6-0.9\,{\rm fm}$ range.
The light blue band represents the ${\rm N^2LO}$ results from the standard
Weinberg approach of Ref.~\cite{Epelbaum:2003xx}.
The dashed dark blue line represents the ${\rm N^2LO}$ results
for $r_c = 0.1\,{\rm fm}$.
}
\label{fig:1S0}
\end{figure}

The results for the singlet $^1S_0$ channel are shown in Fig.~(\ref{fig:1S0}).
Following~\cite{Epelbaum:2003gr,Epelbaum:2003xx},
we take $f_{\pi} = 92.4\,{\rm MeV}$, $m_{\pi} = 138.03\,{\rm MeV}$, 
$g_A = 1.26$, and $d_{18} = -0.97\,{\rm GeV}^2$. 
For the chiral couplings we employ the customary values
$c_1 = -0.81\,{\rm GeV}^{-1}$, $c_3 = -3.40\,{\rm GeV}^{-1}$
and $c_4 = 3.40\,{\rm GeV}^{-1}$,
which are compatible with the determination of Ref.~\cite{Buettiker:1999ap}.
The potential is taken from Ref.~\cite{Rentmeester:1999vw}.
As can be seen, the results reproduce the $^1S_0$ phase shifts
up to $k \sim 200-300\,{\rm MeV}$, depending on the value of the cut-off.
If the cut-off is small ($r_c = 0.1\,{\rm fm}$),
the perturbative treatment of the subleading pieces of the interaction 
starts to fail already at $k \sim 200\,{\rm MeV}$,
as a consequence of the relative weakness of OPE with respect
to the enormous strength of TPE at short distances.
The previous problems can be circumvented by using
cut-offs of the order of $r_c \sim 1 / 2 m_{\pi} \, (0.7\,{\rm fm})$,
which are small enough as to guarantee the correct inclusion of the TPE tail.
In particular we employ $r_c = 0.6-0.9\,{\rm fm}$~\footnote{
Taking into account the relationship
$\Lambda = \pi / 2 r_c$~\cite{Entem:2007jg}, 
the previous configuration space cut-off range is approximately equivalent
to a momentum space (sharp) cut-off of $\Lambda \simeq 350-500\,{\rm MeV}$.}
a range for which perturbative TPE calculations
compete well with non-perturbative ones in the Weinberg
counting at the same order~\cite{Entem:2001cg,Epelbaum:2003xx},
though perturbative TPE is slightly less predictive
due to the additional counterterm.
On a different ground it should be noticed that OPE is perturbative
in the singlet~\cite{Kaplan:1998tg,Kaplan:1998we,Beane:2001bc},
even if iterated~\cite{Fleming:1999ee},
suggesting that the previous results could be reinterpreted
as an ${\rm N^3LO}$/${\rm N^4LO}$ computation
in the KSW counting~\cite{Kaplan:1998tg,Kaplan:1998we}.

The failure of perturbative subleading TPE at $r_c = 0.1\,{\rm fm}$ raises
interesting questions regarding the adequacy of the present power counting
scheme and the role of chiral TPE.
Of course, the technical reasons why perturbation theory fails already
at $k \sim 200\,{\rm fm}$ for small cut-offs are clear:
OPE does not provide enough long range distortion as to avoid higher
momentum waves to probe the van der Waals component of TPE,
as has been discussed for example in Ref.~\cite{Birse:2010jr}.
This component originates from the behaviour of subleading TPE,
which in the singlet channel can be schematically
written as~\cite{Rentmeester:1999vw}
\begin{eqnarray}
2 \mu\,V^{(\nu = 3)}_{\rm TPE}(r) = -
\frac{R_6^4}{r^6}\,e^{-2 m_{\pi} r}\, \sum_{n=0}^{5} a_n (2 m_\pi r)^n \, ,
\end{eqnarray}
where the $a_n$'s are dimensionless parameters with $a_0 = 1$ and
$R_6$ is a length scale related with the strength of
TPE at short distances,
which varies between $R_6 = 1.6-1.8\,{\rm fm}$
for typical values of the chiral couplings.
The previous form implies that the chiral van der Waals component
of subleading TPE should start to become apparent
at distances below $r \leq 1/2m_{\pi} \simeq 0.7\,{\rm fm}\,$.
This figure is supported by several renormalized non-perturbative TPE
computations in the singlet~\cite{Valderrama:2005wv,PavonValderrama:2005uj,Valderrama:2008kj},
which usually reach cut-off independence at distances around or below
$0.5\,{\rm fm}$, signalling the onset of chiral van der Waals forces.
For such cut-off radii the perturbative treatment of TPE generates
terms like $k R_6$ and $m_{\pi} R_6$,
which, taking into account the size of $R_6$,
might cause the perturbative series to eventually diverge.
The most consistent and straightforward solution to this problem 
is the use large enough cut-offs ($r_c > 0.5\,{\rm fm}$)
in order to avoid the conjectured breakdown of
the perturbative series.
The alternative solution, which will not be considered in the present work,
is the iteration of chiral TPE or at least some parts of it~\cite{Birse:2010jr}.
Although interesting, this proposal seems difficult
to harmonize within the EFT framework  
as it requires (i) to justify the promotion of an order $Q^3$ interaction
to order $Q^{-1}$ and (ii) the existence of a cut-off window for which
subleading TPE dominates but the higher order corrections
are still small compared to this contribution.

The employed cut-off window, $r_c = 0.6-0.9\,{\rm fm}$,
represents a compromise between the requirements
of the singlet and triplet channels.
The optimum value of the cut-off in the singlet lies in the vicinity
of $r_c = 0.9-1.0\,{\rm fm}$, a range for which the description
of the triplet phases starts to worsen.
This cut-off window may look soft, but it is not: the first
deeply bound state (i.e. the first zero of the $k=0$ wave function)
for the ${\rm N^2LO}$ potential happens
at $r_c = 0.70\,{\rm fm}$,
meaning that the lower range of the present cut-off window is
already beyond what can be reached in the Weinberg scheme.
It is interesting to notice that the previous cut-off range is similar
to the radii at which most potential models of the NN interaction~\cite{Stoks:1994wp,Wiringa:1994wb,Machleidt:2000ge} 
have their minima, usually at $r \sim 0.8-0.9\,{\rm fm}$.
The mimima mark the distance at which the short range repulsion starts
to overcome the long range attraction, and consequently
can be understood as the separation point
between short ($r \lesssim 0.5\,{\rm fm}$) and
long range ($r \gtrsim 1.0\,{\rm fm}$) physics.
In this sense, the cut-off is to be interpreted as a separation scale,
as has been proposed within the context of
RGA~\cite{Birse:1998dk,Barford:2002je,Birse:2005um},
rather than as a hard scale~\cite{Epelbaum:2006pt,Epelbaum:2009sd}.

In the calculations of Fig.~(\ref{fig:1S0})
we also interpret the cut-off variation of the results
as the error band of the theory.
The previous is a sensible prospect in the sense that we expect the cut-off
dependence of the scattering amplitudes to be a higher order effect.
However, if the cut-off variation is to be understood as an error band,
the size of the band should decrease at each new order
to reflect the convergence properties of the theory.
Paradoxically the ${\rm N^2LO}$ band is bigger than the ${\rm NLO}$ one,
a worrisome situation which does not necessarily mean that we should
abandon the previous interpretation.
In fact the same happens in the Weinberg counting, as illustrated by
the singlet channel results of Ref.~\cite{Epelbaum:2003xx}.
The explanation is to be found in the
surprisingly large size of the $c_3$ and $c_4$ chiral couplings,
which causes the subleading TPE contribution to the chiral potential
to be substantially bigger than the corresponding one from leading TPE.
This is due to the large contributions from the $\Delta$ resonance
to the chiral couplings~\cite{Bernard:1996gq},
$c_{3,\Delta} = - 2 c_{4,\Delta} = - 4 h_A^2 / 9 \Delta$, with $\Delta$
the nucleon-delta mass splitting and $h_A$ the $\pi N \Delta$ axial
coupling, ranging from $-1.7$ to $-2.7\,{\rm GeV}^{-1}$
depending on the value of $h_A$~\footnote{
Taking $h_A$ between $1.08$ and the SU(4) value $1.34$,
see~\cite{Krebs:2007rh}.
The values for the chiral couplings once the $\Delta$ has been included
can also be consulted in Ref.~\cite{Krebs:2007rh}.
}. 
In this sense, the increased size of the ${\rm N^2LO}$ error bands
is just a reflection of the unexpected contribution
from this low energy scale.
The explicit inclusion of the $\Delta$ resonance in the NN chiral potential,
a theme which has been recurrently considered in the literature~\cite{Ordonez:1995rz,Kaiser:1998wa,Krebs:2007rh,Entem:2007jg,Valderrama:2008kj,Valderrama:2010fb},
is presumed to solve the current issue with the error bands
(see also the related discussion of Ref.~\cite{Birse:2010jr}).
This prospect does not appear to be unreasonable in view 
of the perturbative peripheral wave ${\rm N^2LO}$-$\Delta$
results of Ref.~\cite{Krebs:2007rh} and
the related non-perturbative central and peripheral wave calculations
of Refs.~\cite{Valderrama:2008kj,Valderrama:2010fb},
all of which indicate an enhancement in the convergence rate of
the phase shifts as compared to the $\Delta$-less theory.

\section{Triplet Channel}
\label{sec:triplet}

In the case of the $^3S_1-{}^3D_1$ channel the perturbative analysis
is analogous to the previous one for the $^1S_0$ channel,
but more cumbersome due to the presence of coupled channels and
the singular behaviour of the tensor piece of the ${\rm LO}$ potential
in the triplet channels.
The details of such analysis are shown in Appendix~\ref{app:DWBA-triplet},
but the essential point is that the inverse power law
behaviour of the OPE tensor force ($\sim 1/r^3$) softens
the perturbative integrals and reduce the necessary number
of counterterms per phase.
In fact we have that the s- and d-wave wave functions behave as
${{u}_k^{(0)}\,},{{w}_k^{(0)}\,} \sim r^{3/4}$
near the origin~\cite{PavonValderrama:2005gu},
and that each subtraction adds an $r^{5/2}$ suppression
to the short distance behaviour~\footnote{
This is to be compared with the singlet channel, where
${\tilde{u}_k^{(0)}\,}^2 \sim 1$ and each subtraction adds
an additional $r^2$ suppression.
If the singlet ${\rm LO}$ potential had behaved as expected by power
counting, i.e. $1/r^3$, it would have only needed two counterterms at
${\rm NLO}$/${\rm N^2LO}$, following Weinberg's counting.}.
This translates into two subtractions for each of the three phases
in the  $^3S_1-{}^3D_1$ channel ($\delta_{^3S_1}$,$\epsilon_{1}$,
$\delta_{^3D_1}$),
meaning that we end up with six counterterms
at ${\rm NLO}$/${\rm N^2LO}$
in agreement with Ref.~\cite{Valderrama:2005wv}.
That is, the scattering amplitude
can be completely determined using six pieces of data,
for example the value of the three phase shifts at two different momenta.

\begin{figure}[ttt!]
\begin{center}
\epsfig{figure=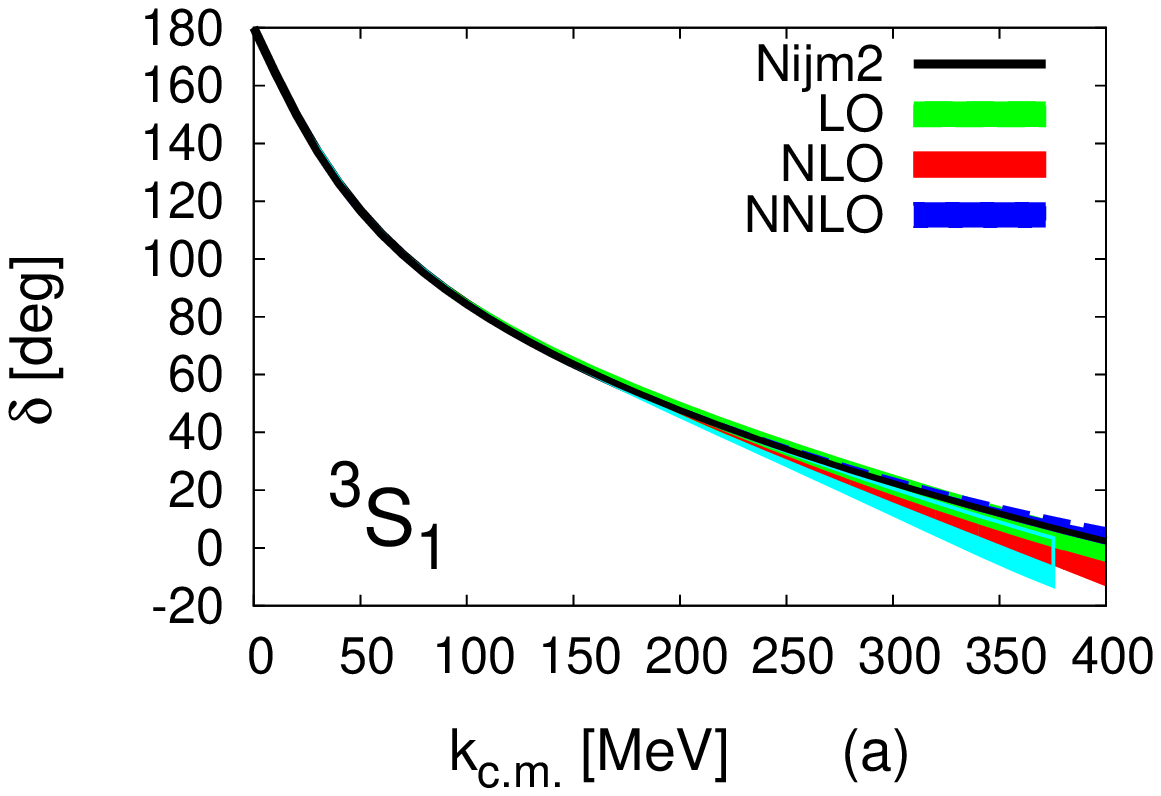, 
	height=5.25cm, width=8.0cm}
\epsfig{figure=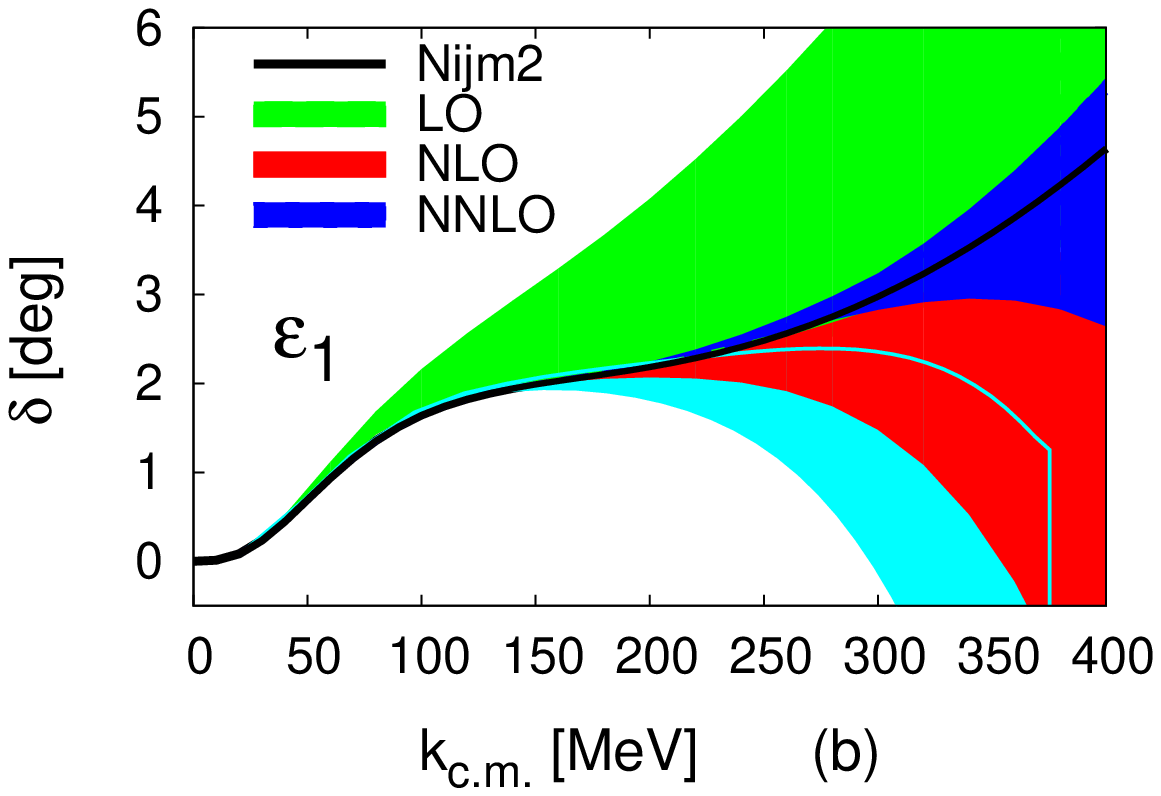, 
	height=5.25cm, width=8.0cm}
\epsfig{figure=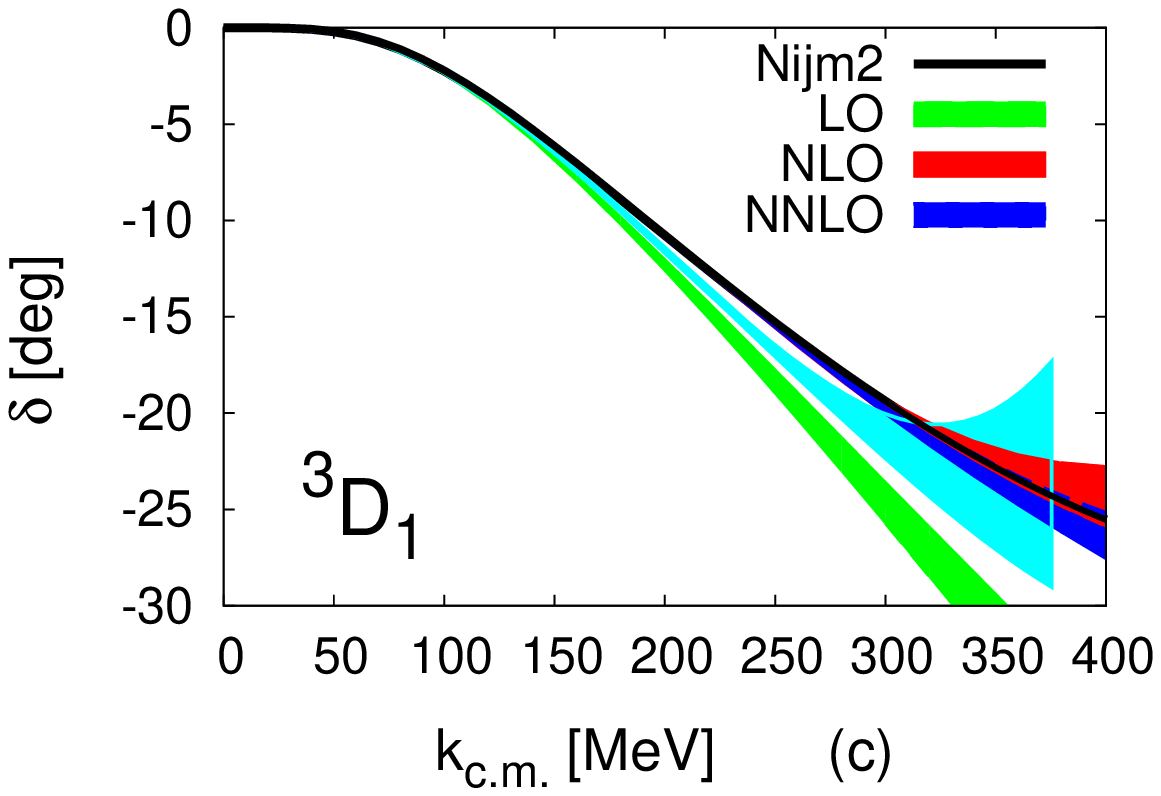, 
	height=5.25cm, width=8.0cm}
\end{center}
\caption{(Color online)
Phase shifts (nuclear bar) for the $^3S_1-{}^3D_1$ coupled channel,
with the ${\rm LO}$ piece (OPE) fully iterated and the ${\rm NLO}$ and 
${\rm N^2LO}$ pieces (chiral TPE) treated perturbatively.
The ${\rm LO}$ counterterm is fixed to reproduce
the triplet scattering length $a_{0,t} = 5.419\,{\rm fm}$.
The error bands and the fitting range are the same as in Fig.~(\ref{fig:1S0}).
The light blue band corresponds to the ${\rm N^2LO}$ results
of Ref.~\cite{Epelbaum:2003xx} in the standard Weinberg counting.
The dashed dark blue line represents the ${\rm N^2LO}$ results for 
$r_c = 0.3\,{\rm fm}$.
}
\label{fig:3C1}
\end{figure}

The results are shown in Fig.~(\ref{fig:3C1}).
Perturbative TPE provides a good description of
the $^3S_1$ and $^3D_1$ phases and the $\epsilon_1$ mixing angle 
up to moderately high momenta, around $k \sim 300-400\,{\rm MeV}$,
although it should be noted that the results are quite sensitive to the choice
of chiral couplings, due to the linear dependence generated by
treating chiral TPE perturbatively.
Contrary to the singlet case, small cut-offs do not affect the
momentum range in which first order perturbation theory works,
although due to numerical limitations, the cut-off cannot be
reliably reduced below $r_c = 0.3\,{\rm fm}$.
However, there are reasons for keeping the cut-off in the proposed
window, such as avoiding unphysical deeply bound states in the leading order
amplitudes (the first one appears at $r_c = 0.45\,{\rm fm}$),
or an excessive D-state probability in the deuteron,
yielding poor convergence in nuclear matter
calculations~\cite{Machleidt:2009bh}.
Larger cut-offs, of the order of $1\,{\rm fm}$ and above, are also
disfavoured as they lead to a worse description of the $\epsilon_1$
mixing angle for momenta above $300\,{\rm MeV}$,
similar to the one obtained
in the ${\rm N^2LO}$ Weinberg calculation of Ref.~\cite{Epelbaum:2003xx}.
The proposed cut-off range avoids the previous problems and,
due to the stronger long range distortion provided
by the tensor component of OPE,
generate error bands which decrease in size order by order.

In this regard, it is interesting to notice the opposite cut-off
preferences of the singlet and triplet channels. 
The mismatch in the preferred cut-off windows is a reflection of
the different physics at play in these waves.
In the singlet, all pion exchanges are perturbative and
the iteration of OPE is merely a short-cut to avoid
the computation of higher order perturbations,
while in the triplet tensor OPE really needs to be iterated.
Different power countings require different cut-off windows.
In this sense, large cut-off values worsen the convergence
of the triplet: the OPE tensor force starts to behave perturbatively,
even if fully iterated in the Schr\"odinger equation.
This entails a change in the counting of the triplet channel
from the modified Weinberg scheme of Nogga, Timmermans
and van Kolck~\cite{Nogga:2005hy}
eventually to KSW~\cite{Kaplan:1998tg,Kaplan:1998we},
thus reducing the convergence of the theory as the cut-off is increased.
The exact point at which the change takes place is difficult to determine,
but probably lies above $r_c \gtrsim 1/m_{\pi} = 1.4\,{\rm fm}$.
Of course, the fact that the $\epsilon_1$ mixing angle is the phase which
starts to feel the problem earlier is not a surprise, as it depends on
delicate cancellations between short and long range effects. 
The singlet channel, on the contrary, does not have any problem with larger
values of the cut-off as the power counting is not changed:
the use of larger cut-off values only entails a rearrangement of
the short range physics to account for those parts of the pion tail
which have been ignored, but the assumption
that all long range interactions are perturbative remains unchanged.

\section{The Role of the Cut-off}
\label{sec:cut-off}

In the previous calculations we have taken a very pragmatic
point of view with regard to the cut-off:
we have chosen the cut-off range $r_c = 0.6-0.9\,{\rm fm}$ in order
to improve the convergence of the theory and the description of
the phase shifts up to ${\rm N^2LO}$.
Of course, the choice of this range depends on a compromise between
the specific requirements of the singlet and triplet channels,
as explained in the previous sections.
The important point is however that
the proposed cut-off window generates leading order phase shifts
which do not differ too much from the Nijmegen ones,
an arrangement which minimizes the size of the subleading order corrections
and, as a consequence, enhances the convergence of the theory.

This criterion basically coincides with the interesting cut-off philosophy
of Beane, Kaplan and Vuorinen~\cite{Beane:2008bt},
in which the cut-off is merely a parameter controlling
the convergence rate of the theory.
The underlying idea behind this interpretation is an analogy with
the role of renormalization scale dependence in QCD (see
for example~\cite{PhysRevD.23.2916,PhysRevD.28.228}).
A similar rationale can be provided by the observation that
the full scattering amplitude, computed at all orders,
is cut-off independent as a consequence of having
an infinite number of counterterms.
In this regard, cut-off dependence is just an artifact
of finite order approximations
which can be avoided by the careful selection of 
a cut-off window
for which the particular power counting under consideration is realized.

However, for this interpretation to be complete within an EFT context
it is necessary to determine first some formal aspects of the theory,
such as the expansion parameter and the cut-off and momentum ranges
for which the perturbative expansion converges.
The knowledge of the expansion parameter is fundamental in order
to be able to make rigorous error estimations of the results
and to check the suitability of the selected cut-off window.
On the other hand, the determination of the range of applicability of the EFT
is necessary for avoiding power counting abuse, that is, 
claiming as legitimate the accidental description of data
beyond the possibilities of the EFT under consideration.

Due to the mostly numerical nature of the present investigation,
it is not clear how to extract the expansion parameter.
However, the deconstruction of the $^1S_0$ singlet channel of Ref.~\cite{Birse:2010jr}
might provide some valuable clues regarding this important aspect of
the theory.
The energy dependence of the short range physics in this channel
suggests a breakdown scale of $\Lambda_{0,s} \simeq 270\,{\rm MeV}$.
This estimation translates into an expansion parameter of
$m_{\pi} /  \Lambda_{0,s} \simeq 0.5$, a value which is compatible
with the conjectured equivalence of the present approach
with the KSW counting in the singlet channel.
For the $^3S_1-{}^3D_1$ triplet channel there is no deconstruction yet
which might provide a preliminary estimation of the breakdown scale,
but if we assume the deconstruction of the p-wave uncoupled
triplets~\cite{Birse:2007sx}
to hold in the $^3S_1-{}^3D_1$ coupled triplet,
we obtain $\Lambda_{0,t} \simeq 340\,{\rm MeV}$~\footnote{
This corresponds to a laboratory energy of $250\,{\rm MeV}$, above which
the short range interaction in the $^3P_0$, $^3P_1$ and $^3D_2$ waves
cannot be reliably described by two counterterms~\cite{Birse:2007sx}.
On the other hand, the assumption that the p-waves yield a good approximation
for the breakdown scale of the s-waves is not unreasonable
if we take into account that the deconstruction
of the $^1P_1$ wave~\cite{Ipson:2010ah} basically suggests
the same estimation as the $^1S_0$ wave~\cite{Birse:2010jr}.}.
The related expansion parameter would be $m_{\pi} /  \Lambda_{0,t} \simeq 0.4$,
a value which is compatible with the observation that the convergence
in the triplet is better than in the singlet.

\section{Relation to Other Approaches}
\label{sec:approaches}

In the present work we determine the power counting of the counterterms
by requiring the renormalizability of the perturbative
corrections to the scattering amplitude,
where by renormalizability it is understood the elimination of
all negative powers of the coordinate space cut-off $r_c$.
There is still a residual cut-off dependence which is nominally of
higher order, meaning that perturbative renormalizability implies
the cut-off independence of the scattering amplitude
at the order considered.

This is very similar to the RG approach of Birse~\cite{Birse:1998dk},
where the relative scaling (i.e. the power counting) of the counterterms
is determined by requiring the cut-off independence
of the scattering amplitude.
Of course, exact cut-off independence is only achieved at infinite order.
Finite order truncations will lead to a residual cut-off dependence involving
positive powers of the cut-off $r_c$, but the renormalizability of the
amplitudes is guaranteed.
Therefore, it is not surprising a great degree of agreement between these
two approaches.

This expectation is realized in the singlet channel,
where renormalization group analysis~\cite{Birse:2005um}
and deconstruction~\cite{Birse:2010jr}
are equivalent to perturbative renormalizability.
For the triplet channel the situation is mixed:
in the case of the $\epsilon_1$ mixing angle and the $^3D_1$ phase,
the observation that two counterterms are needed to renormalize
each of these phases is compatible with the deconstruction
of the p- and d-wave uncoupled triplets of Ref.~\cite{Birse:2007sx}.
However, the RGA of Ref.~\cite{Birse:2005um} predicts
one additional counterterm for the $^3S_1$ phase
which should appear at order $Q^{5/2}$.
This counterterm is not needed by perturbative renormalizability.

The previous discrepancy is surprising: we are making the same assumptions
as Ref.~\cite{Birse:2005um} regarding which pieces of the interaction
to iterate, yet the resulting power countings are slightly different.
However, this is not new: 
the non-perturbative renormalizability of the OPE potential
dictates that each attractive triplet requires one
counterterm, while repulsive triplets do not.
On the contrary, the RGA of Ref.~\cite{Birse:2005um} makes no distinction
for the power counting of attractive and repulsive triplets.
The paradigmatic example is given by the $^3P_0$ (attractive)
and $^3P_1$ (repulsive) waves.
As happened in the peripheral waves, the inconsistency can be circumvented
in terms of the perturbative analysis of tensor OPE of Ref.~\cite{Birse:2005um}:
for the $^3P_0$ wave the perturbative treatment of OPE is expected to fail
already at $k\sim 200\,{\rm MeV}$,
while for the $^3P_1$ wave this limit is extended up to $k \sim 400\,{\rm MeV}$.
Therefore, in the range of momenta of interest for nuclear EFT
the $^3P_1$ wave can in principle be described
in terms of the original Weinberg counting~\footnote{
One could think of extending this argument to the $^3D_1$ phase, 
which is usually well reproduced in perturbation theory~\cite{Kaiser:1997mw}.
However, taking into account the coupled channel nature of the $^3D_1$ phase,
it is probably inconsistent to treat tensor OPE perturbatively
in the d-wave channel but not in the s-wave channel.
}.

For the $^3S_1$ phase the causes of the disagreement are to be found
in the naive extrapolation of the idea of trivial and
non-trivial fixed points to a problem where
these concepts may not be applicable.
The relevant observation in this context
is that attractive singular potentials
do not have a unique solution~\cite{Beane:2000wh,Valderrama:2005wv,PavonValderrama:2005uj}:
the value of the scattering
length oscillates indefinitely as the cut-off varies,
a situation which is solved by the inclusion
of a counterterm, stabilizing the solution.
In this regard, for an attractive singular interaction all values of the
scattering length are equally fine-tuned, implying that the distinction
between trivial and non-trivial fixed points is artificial in this case.
As analyzed in Ref.~\cite{PavonValderrama:2007nu},
the RG evolution of attractive singular potentials
is driven in the infrared limit~\footnote{Notice however that
Ref.~\cite{PavonValderrama:2007nu} uses a different language
than Birse's RGA~\cite{Birse:2005um}:
what is called ultraviolet (long range) limit in~\cite{PavonValderrama:2007nu}
corresponds to the infrared (short range) limit of~\cite{Birse:2005um}.
If we call the light and heavy scales $m_l$ and $m_h$,
Ref.~\cite{PavonValderrama:2007nu} is taking $m_l r_c \to 0$,
while Ref.~\cite{Birse:2005um} assumes $m_h \gg 1/r_c \gg m_l$,
or equivalently $m_h r_c \to \infty$, $m_l r_c \to 0$ and
$m_l / m_h \to 0$.
Contrary to Ref.~\cite{Birse:2005um}, Ref.~\cite{PavonValderrama:2007nu}
does not analyze the power counting of the short range operators
but rather concentrates on issues such as the cut-off dependence
of observables and the fixed points, limit cycles and attractors
which result from the RG flow of regular and singular potentials. 
}
towards an oscillatory attractor-type solution resembling a limit cycle.
However, the attractor-type solution does not have the discrete scaling
properties of limit cycles (see~\cite{PavonValderrama:2007nu} for details).

Alternatively, the previous observations can also be understood in terms
of the behaviour of the squared reduced wave functions
at short distances.
For regular potentials there are two possible behaviours, the regular
one, $|u(r_c)|^2 \sim r_c^2$ ,which can be identified
with ``natural'' systems, and the irregular one, $|u(r_c)|^2 \sim 1$,
which describes systems with unnaturally large scattering lengths.
On the contrary, for attractive singular potential, the wave function
always behaves as $|u(r_c)|^2 \sim r_c^{3/2}$ (times an oscillatory factor),
independently of the value of the scattering length.
That is, there is no additional short range enhancement due to
large scattering lengths.
In this regard, we should not expect the existence of
two different kinds of fixed points in the RG flow
of attractive singular interactions.
The previous observations indicate that for attractive singular potentials
(i) the $C_0$ counterterm must be of order $Q^{-1}$, as required by
non-perturbative renormalizability and (ii) the first perturbation
to the $C_0$ counterterms is of order $Q^{-1/2}$ as expected from
the behaviour of the squared wave function, meaning that
the attractor is a stable solution of the RG flow.
Consequently, the correct RG analysis for channels with an attractive tensor
force is the one termed ``trivial'' in Ref.~\cite{Birse:2005um},
conveniently modified to incorporate the previous observation
about the $C_0$ operator.

A recent work which is also relevant for the present approach is
the toy model proposed by Epelbaum and Gegelia to address
the role of regularization and renormalization in EFT~\cite{Epelbaum:2009sd}.
In this work, the authors consider a two-body potential problem which
shares many of the features of nuclear EFT,
like the existence of a separation of scales or the possibility of
expanding the long range interaction in terms of a power counting.
The conclusions of the analysis of Epelbaum and Gegelia support
most of the assumptions usually invoked in the Weinberg scheme,
namely that naive dimensional analysis provides a good enough power
counting and the ideal value of the cut-off should
be chosen of the order of the hard scale of the problem.
In addition, if the cut-off is taken much beyond the hard scale,
the non-perturbative renormalization procedure may break the assumptions
made in the first place by the power counting,
a phenomenon which Epelbaum and Gegelia call ``peratization''.

The lessons derived from a specific toy model may be however of limited
significance.
In particular, there is an essential feature of the chiral expansion
which is not reproduced in the previous model,
namely the appearance of singular interactions
at leading and subleading orders.
Contrary to the expectations of Epelbaum and Gegelia, 
the presence of singular potentials implies that
(i) non-perturbative power counting will break down at cut-offs much softer
than expected and that (ii) deviations from naive dimensional analysis may
eventually happen.
These aspects have been probably overlooked in the previous analysis due to
the very good properties of the toy model:
subleading contributions to the toy potential are only mildly divergent
and, in addition, they are always suppressed by the expected ratio
of low energy versus high energy scales.
On the contrary, the subleading pieces of the chiral NN potential
can receive unexpectedly large contributions
from light degrees of freedom which have not been explicitly
taken into account, like the $\Delta$ resonance.
It is not surprising therefore that a toy model incorporating 
many of our naive expectations about EFT turns out to confirm them.

However, as far as we limit ourselves to soft enough cut-offs,
the conclusions of Epelbaum and Gegelia regarding naive
dimensional analysis (i.e. Weinberg counting) are likely to hold.
This observation is realized
in the work of Shukla et al.~\cite{Shukla:2008sp}
which, much in the spirit of deconstruction, analyzes
the short distance physics of the $^1S_0$ singlet channel
with the chiral NN potential up to ${\rm N^2LO}$.
The authors observe that in the cut-off region $r_c = 1.0-1.8\,{\rm fm}$
two counterterms are enough to parametrize the short range physics,
a finding consistent with the idea that the Weinberg counting
is better realized for soft values of the cut-off.
A particularly interesting aspect of the previous work is
the reanalysis of the short range physics for perturbative chiral TPE.
For the cut-off range $r_c = 1.4-1.8\,{\rm fm}$ the extracted
short range physics can be accurately approximated by
first order perturbative TPE, 
while for the region $r_c = 1.0-1.4\,{\rm fm}$ one needs to go
to second and third order in the perturbative series in order
to reproduce the non-perturbative results,
although there is still convergence.
In the softer cut-off range the Weinberg scheme is perfectly realized
as a perturbative power counting.
For the harder cut-off range, Weinberg is still a consistent (non-perturbative)
power counting scheme, as subleading order corrections are smaller
than leading order ones.
The efforts of Ref.~\cite{Shukla:2008sp} probably represent the best way
to analyze the merits of the Weinberg counting in realistic cases.
The extension to other partial waves, in particular the triplet,
would be very welcomed.

If the cut-off is decreased below $R_0 = 1.0\,{\rm fm}$,
the authors of Ref.~\cite{Shukla:2008sp} observe that
the contributions from subleading TPE start to grow uncontrollably,
signalling the breakdown of the Weinberg counting.
Below this cut-off, power counting is likely to be lost
in non-perturbative calculations, as loop contributions
from the subleading pieces will eventually dominate
the amplitudes.
The previous breakdown scale is however uncomfortably soft:
using the equivalence $\Lambda = \pi / 2 r_c$~\cite{Entem:2007jg},
$R_0$ naively corresponding to a (sharp) momentum
cut-off of $\Lambda_0 \simeq 310\,{\rm MeV}$.
Most Weinberg calculations use momentum space cut-offs of the order of
$\Lambda \sim 0.5\,{\rm GeV}$, which may be hard enough as to peratize
the amplitudes.
As suggested in Ref.~\cite{Valderrama:2010aw},
this may be already happening in the $^1S_0$ singlet channel
for $\Lambda = 400\,{\rm MeV}$ at ${\rm N^2LO}$.
These observations do not imply however that Weinberg counting is not useful,
only that it should be employed within its specific range of applicability.
In this respect,
the most interesting feature of perturbative treatments 
is that they are guaranteed to respect the power counting
independently of the value of the cut-off,
precluding from the start the possibility of any power counting inconsistency.

A recent work which is also relevant for the discussion is
the new KSW expansion of Beane, Kaplan and Vuorinen~\cite{Beane:2008bt},
which challenges one of the key premises of the present approach,
namely  that OPE should be fully iterated in the triplet,
by constructing a viable nuclear EFT in which all pion exchanges
are treated as perturbations.
In this work the convergence problems of
the original KSW counting~\cite{Fleming:1999ee,Birse:2005um}
are alleviated by the exchange of a fictitious meson of mass $\lambda$
which regulates the $1/r^3$ singularity of the tensor force at short distances.
For the optimum value of the regulator ($\lambda = 750\,{\rm MeV}$),
the expansion apparently converges up to order $Q$, albeit slowly.
At this order, the results of Ref.~\cite{Beane:2008bt}
for the $^3S_1$ and $^3D_1$ phases compare well with
the ${\rm LO}$ results of the present approach.
However, the order $Q$ results for the $\epsilon_1$ mixing angle is
clearly worse than our ${\rm LO}$ computation
and it does not seem to converge for $k > m_{\pi}$.
This may be a good indicator that the tensor force really needs to be iterated,
as the $\epsilon_1$ mixing angle is very sensitive to large cancellations
between long and short range physics.
In any case, a serious comparison of the present approach with the proposal of
Beane, Kaplan and Vuorinen~\cite{Beane:2008bt} requires (i)
the extension of the previous results beyond order $Q$ and
(ii) the consideration of the $^3P_0$ phase
which according to Nogga, Timmermans and van Kolck~\cite{Nogga:2005hy}
also demands the non-perturbative inclusion of tensor OPE.

The observation that OPE is perturbative in the singlet and non-perturbative
in the triplet is closely related with the proposal of
Beane, Bedaque, Savage and van Kolck (BBSvK)~\cite{Beane:2001bc},
which suggested the iteration of those pieces of the (leading order)
chiral ${\rm NN}$ potential which survive in the chiral limit
(that is, tensor OPE).
This prescription is theorized to generate a convergent expansion of
the scattering amplitudes around the chiral limit, therefore
providing a consistent EFT expansion
for two-nucleon systems.
The existence of a deeper relationship
with the present approach remains to be seen.
However, the consideration of the subleading orders of the potential can break
the correspondence,
as there are pieces of these contributions to the potential
which survive in the chiral limit and which are strong enough as
to be iterated, particularly in the singlet channel.
In this regard, the BBSvK scheme might provide a justification
for the iteration of chiral van der Waals forces.

\section{Conclusions}
\label{sec:conclusions}

The present approach determines the momentum and cut-off range
for which chiral TPE behaves perturbatively
when renormalizability is imposed.
The use of small cut-offs is straightforward, but reduces the range of
applicability of the theory in the singlet channel.
The calculations turn out to confirm the viability of the counting proposal of
Nogga, Timmermans and van Kolck~\cite{Nogga:2005hy},
and corroborate to a large extent the related RGA by Birse~\cite{Birse:2005um},
which predicted the power counting of the short range operators.
There are some minor discrepancies however between perturbative
renormalizability and RGA in the triplet channel, 
specifically for the $^3S_1$ phase, which are understood,
suggesting minor modifications and possible improvements
to the RGA of~\cite{Birse:2005um}.
However there are some formal aspects of the present EFT formulation
which need to be elucidated, like the role of the cut-off,
the determination of the expansion parameter or
the range of applicability of perturbative TPE.
In this regard, the deconstruction approach of Refs.~\cite{Birse:2007sx,Birse:2010jr,Ipson:2010ah}
is able to provide some interesting clues and preliminary answers.
Of course, a complete evaluation of the renormalized perturbative treatment
of chiral TPE should also include the calculation of the p- and d-wave
phase shifts and the deuteron properties.
The present analysis paves the way for such computations, which we leave for
future works. 

\begin{acknowledgments}

I would like to thank E.~Epelbaum,
A.~Nogga and E. Ruiz Arriola for discussions
and a critical and careful reading of the manuscript.
I would also like to thank E.~Epelbaum for kindly providing the data
corresponding to the ${\rm N^2LO}$ phase shifts of Ref.~\cite{Epelbaum:2003xx},
D.R.~Phillips for encouragement and discussions
and M.C.~Birse for discussions.
This work was supported by the Helmholtz Association fund
provided to the young investigator group
``Few-Nucleon Systems in Chiral Effective Field Theory'' (grant
VH-NG-222), the virtual institute ``Spin and strong QCD'' (VH-VI-231),
the EU Research Infrastructure Integrating Initiative HadronPhysics2
and the the Spanish Ingenio-Consolider 2010 Program CPAN (CSD2007-00042).

\end{acknowledgments}

\appendix

\section{Derivation of the DWBA for the Singlet Channel}
\label{app:DWBA-singlet}

In this appendix, we derive the DBWA formulae used along the present paper.
We start by considering a potential which can be decomposed as a zeroth order
approximation and a perturbation
\begin{eqnarray}
V(r) = V^{(0)}(r) + V^{(1)}(r)\, ,
\end{eqnarray}
and the related reduced Schr\"odinger equations for the zeroth order and 
full reduced wave functions, $u_k^{(0)}$ and $u_k$
\begin{eqnarray}
- {u_k^{(0)}}'' + 2\mu\,V^{(0)}\,u_k^{(0)} &=& k^2\,u_k^{(0)} \, , \\
- {u_k}'' + 2\mu\,[ V^{(0)} + V^{(1)} ]\,u_k &=&  k^2\,u_k\, ,
\end{eqnarray}
where $\mu$ is the reduced mass of the system.
The full reduced wave function can be perturbatively expanded as
\begin{eqnarray}
u_k^{(0+1)}(r) = u_k^{(0)}(r) + u_k^{(1)}(r) + {\mathcal O}({(V^{(1)})}^2) \, ,
\end{eqnarray}
where, for the purposes of this work, it is enough to consider first order
perturbation theory only.

In order to obtain the DWBA expressions we begin by
(i) multiplying the zeroth order Schr\"odinger equation by the full solution
$u_k$, and (ii) the full Schr\"odinger equation by the zeroth order
solution $u_k^{(0)}$.
Then we compute the difference between (i) and (ii), yielding
\begin{eqnarray}
{\left( u_k^{(0)}\,u_k'  -  {u_k^{(0)}}'\,u_k \right)}' =
2\mu\,V^{(1)}(r)\,u_k^{(0)}(r)\,u_k(r) \, . \nonumber\\
\end{eqnarray}
The expression above can be integrated to obtain the Wronskian identity
\begin{eqnarray}
W ( u_k^{(0)}, u_k) {\Big|}_{r_c}^{R} =
2\mu\,\int_{r_c}^R\,dr\,V^{(1)}(r)\,u_k^{(0)}(r)\,u_k(r) \, ,
\nonumber \\
\end{eqnarray}
where $W(f,g) = f(r) g'(r) - f'(r) g(r)$ is the Wronskian,
and $r_c$ and $R$ are respectively the ultraviolet and
infrared cutoffs.
The infrared cutoff $R$ can be eliminated  by taking into account
the long distance behaviour of the $u_k^{(0)}$ and $u_k$ reduced
wave functions, which is given by
\begin{eqnarray}
u_k^{(0)}(r) &\stackrel{r\to\infty}{\longrightarrow}& 
\frac{1}{\mathcal{A}^{(0)}(k)}\,
\frac{\sin{(k\,r + \delta^{(0)})}}{\sin{\delta^{(0)}}} \, , \\
u_k(r) &\stackrel{r\to\infty}{\longrightarrow}& 
\frac{1}{\mathcal{A}(k)}\,
\frac{\sin{(k\,r + \delta)}}{\sin{\delta}} \, ,
\end{eqnarray}
where $\mathcal{A}^{(0)}(k)$ and $\mathcal{A}(k)$ are normalization factors
which ensure an energy independent normalization of the reduced wave
functions at the cut-off radius.
With the previous wave functions the Wronskian can be evaluated at 
$R \to \infty$, resulting in
\begin{eqnarray}
W ( u_k^{(0)}, {u_k^{(0+1)}}) {\Big|}_{R} = 
- \frac{k}{\mathcal{A}^{(0)}\,\mathcal{A}}\,
\frac{\sin{(\delta - \delta^{(0)})}}{\sin{\delta}\,\sin{\delta^{(0)}}} \, .
\end{eqnarray}
Therefore, we arrive at the following expression
\begin{eqnarray}
&& \frac{k}{\mathcal{A}^{(0)}\,\mathcal{A}}\,
\frac{\sin{(\delta - \delta^{(0)})}}{\sin{\delta}\,\sin{\delta^{(0)}}} +
f(r_c) = \nonumber \\ && \quad
-2\mu\,\int_{r_c}^{\infty}\,dr\,V^{(1)}(r)\,u_k^{(0)}(r)\,u_k(r) \, ,
\end{eqnarray}
where $f(r_c)$ is just the Wronskian evaluated at $r = r_c$, i.e.
$f(r_c) = W ( u_k^{(0)}, u_k) {|}_{r_c}$,
which does not depend on the momentum $k$ as a consequence of the energy
independent normalization at $r = r_c$.
The perturbative expansion of the previous formula can be obtained from
the corresponding one of its components
\begin{eqnarray}
\delta(k) &=& \delta^{(0)}(k) + \delta^{(1)}(k) + 
{\mathcal O}({(V^{(1)})}^2) \, , \\
u_k(r) &=& u_k^{(0)}(r) + u_k^{(1)}(r) + {\mathcal O}({(V^{(1)})}^2) \, , \\
\mathcal{A}(k) &=& \mathcal{A}^{(0)}(k) + \mathcal{A}^{(1)}(k) +
{\mathcal O}({(V^{(1)})}^2) \, , \\
f(r_c) &=& f^{(1)}(r_c) + {\mathcal O}({(V^{(1)})}^2) \, ,
\end{eqnarray}
yielding the following DWBA formula for the phase shift
\begin{eqnarray}
&& \frac{k}{{\mathcal{A}^{(0)}\,}^2}\,
\frac{\delta^{(1)}(k; r_c)}{\sin{\delta^{(0)}}^2} +
f^{(1)}(r_c) = \nonumber \\ && \quad
-2\mu\,\int_{r_c}^{\infty}\,dr\,V^{(1)}(r)\,{u_k^{(0)}\,}^2(r) \, ,
\end{eqnarray}
where the Wronskian term $f^{(1)}$
can be safely ignored in renormalized computations,
as it vanishes once the first subtraction is done.

\section{DWBA for the Triplet Channel}
\label{app:DWBA-triplet}

In this appendix we present the perturbative distorted wave formulas
for the phase shifts in the $^3S_1-{}^3D_1$ triplet channel.
For that, we express the phase shifts as the expansion 
\begin{eqnarray}
\delta_{\alpha}(k; r_c) &=&
\delta^{(0)}_{\alpha} + \delta^{(2)}_{\alpha} + 
\delta^{(3)}_{\alpha} + \mathcal{O}(Q^4) \, , \\
\delta_{\beta}(k; r_c) &=&
\delta^{(0)}_{\beta} + \delta^{(2)}_{\beta} + 
\delta^{(3)}_{\beta} + \mathcal{O}(Q^4) \, , \\
\epsilon (k; r_c) &=&
\epsilon^{(0)} + \epsilon^{(2)} + 
\epsilon^{(3)} + \mathcal{O}(Q^4) \, ,
\end{eqnarray}
where we have chosen the eigen parametrization of the phase shifts~\cite{PhysRev.86.399}
because in this parametrization the DWBA formulas take their simplest form.
The expansion of the nuclear bar phase shifts~\cite{PhysRev.105.302}
can be obtained from the previous one by reexpanding the relationships
\begin{eqnarray}
\bar{\delta}_1 + \bar{\delta}_2 &=& {\delta}_{\alpha} + {\delta}_{\beta} \, , \\
\sin{( \bar{\delta}_{1} - \bar{\delta}_{2} )} &=&  
\frac{\tan{2\bar{\epsilon}}}{\tan{2\epsilon}} \, , \\
\sin{( {\delta}_{\alpha} - {\delta}_{\beta} )} &=& 
\frac{\sin{2\bar{\epsilon}}}{\sin{2\epsilon}} \, ,
\end{eqnarray}
according to the counting.
The LO phase shifts, $\delta_{\alpha}^{(0)}$, $\delta_{\beta}^{(0)}$ and
$\epsilon^{(0)}$, are obtained by solving non-perturbatively
the OPE potential with one counterterm, which is used for
fixing the triplet scattering length to $a_t = 5.419\,{\rm fm}$.
The exact procedure is explained in Ref.~\cite{PavonValderrama:2005gu}.
The expressions for the perturbative corrections to the LO phase shifts
are the following
\begin{eqnarray}
\label{eq:dnu_alpha_pert_unreg}
\frac{\delta_{\alpha}^{(\nu)}(k; r_c)}{\sin^2{\delta_{\alpha}^{(0)}}} &=& - 
\frac{2\mu}{k}\,{\mathcal{A}^{(0)}_{\alpha}}^2(k)\,
I_{\alpha \alpha}^{(\nu)} (k; r_c) \, , \\ 
\frac{\delta_{\beta}^{(\nu)}(k; r_c)}{\sin^2{\delta_{\beta}^{(0)}}} &=& - 
\frac{2\mu}{k^5}\,{\mathcal{A}^{(0)}_{\beta}}^2(k)\,
I_{\beta \beta}^{(\nu)} (k; r_c) \, , \\
\epsilon^{(\nu)}(k; r_c) &=& - \frac{2\mu}{k^3}\,
\frac{{\mathcal{A}^{(0)}_{\beta}}(k)\,{\mathcal{A}^{(0)}_{\alpha}}(k)}
{\cot{\delta^{(0)}_{\beta}} -\cot{\delta^{(0)}_{\alpha}}}\,
I^{(\nu)}_{\beta \alpha}(k; r_c) \, , \nonumber \\
\end{eqnarray}
where the perturbative integrals $I_{\alpha \alpha}$, $I_{\beta \alpha}$ and
$I_{\beta \beta}$ are defined as
\begin{eqnarray}
I^{(\nu)}_{\rho \sigma}(k; r_c) &=&
\int_{r_c}^{\infty}\,dr\,\Big[ V_{\rm ss}^{(\nu)}(r)\,
{u}_{k,\rho}^{(0)}\,(r)\,{u}_{k,\sigma}^{(0)}\,(r) + 
\nonumber \\ && V_{\rm sd}^{(\nu)}(r)\,\Big( 
{u}_{k,\rho}^{(0)}\,(r)\,{w}_{k,\sigma}^{(0)}\,(r) + \nonumber \\
&& \quad \quad \quad \quad
{w}_{k,\rho}^{(0)}\,(r)\,{u}_{k,\sigma}^{(0)}\,(r)
\Big) + 
\nonumber \\ && V_{\rm dd}^{(\nu)}(r)\,
{w}_{k,\rho}^{(0)}\,(r)\,{w}_{k,\sigma}^{(0)}\,(r) \Big] \, , 
\end{eqnarray}
with $\rho,\sigma = \alpha, \beta$.
As in the singlet case,
$\mu$ represents the reduced mass of the system,
$u_{k,\alpha(\beta)}^{(0)}$ and $w_{k,\alpha(\beta)}^{(0)}$
are the leading order s- and d-wave reduced wave functions
for the $\alpha$($\beta$) scattering states in an energy
independent normalization at the origin / cut-off radius,
and ${\mathcal{A}^{(0)}_{\alpha}}$ and ${\mathcal{A}^{(0)}_{\beta}}$ are
the normalization factors which ensure the previous condition.
The asymptotic normalization of the $\alpha$ and $\beta$ scattering states
is taken to be
\begin{eqnarray}
{\mathcal{A}^{(0)}_{\alpha}}\,u_{k,\alpha}^{(0)}(r) &\to& \cos{\epsilon^{(0)}}\,
( \cot{\delta_{\alpha}^{(0)}}\,\hat{j}_0 (k r) - \hat{y}_0 (k r) )
\, ,\nonumber \\
{\mathcal{A}^{(0)}_{\alpha}}\,w_{k,\alpha}^{(0)}(r) &\to& \sin{\epsilon^{(0)}}\,
( \cot{\delta_{\alpha}^{(0)}}\,\hat{j}_2 (k r) - \hat{y}_2 (k r) )
\, , \nonumber \\
\\
k^2\,{\mathcal{A}^{(0)}_{\beta}}\,u_{k,\beta}^{(0)}(r)
&\to& -\sin{\epsilon^{(0)}}\,
( \cot{\delta_{\beta}^{(0)}}\,\hat{j}_0 (k r) - \hat{y}_0 (k r) )
\, ,\nonumber \\
k^2\,{\mathcal{A}^{(0)}_{\beta}}\,w_{k,\beta}^{(0)}(r) &\to& 
\phantom{-}\cos{\epsilon^{(0)}}\,
( \cot{\delta_{\beta}^{(0)}}\,\hat{j}_2 (k r) - \hat{y}_2 (k r) )
\, , \nonumber \\
\end{eqnarray}
where $\hat{j}_l(x) = x j_l(x)$, $\hat{y}_l(x) = x y_l(x)$,
with $j_l(x)$ and $y_l(x)$ the spherical Bessel functions.
Due to the energy independent normalization of the wave functions
at the cut-off radius, they can be expanded at short distances
as~\cite{PavonValderrama:2005gu}
\begin{eqnarray}
u^{(0)}_{k,\alpha(\beta)}(r) &=&
\sum_{n=0}^{\infty} u^{(0)}_{2n,\alpha(\beta)}(r) k^{2n} \, ,
\end{eqnarray}
where the behaviour is given by
\begin{eqnarray}
u^{(0)}_{2n,\alpha(\beta)}(r) \sim r^{3/4 + 5n/2}\,f(\sqrt{\frac{a}{r}}) \, ,
\end{eqnarray}
with $f(x)$ some combination of $\sin{x}$, $\cos{x}$ and $e^{-\sqrt{2}x}$.
The length scale $a$  is related to the strength of the tensor force.
In principle the general solution of the Schr\"odinger equation
for the tensor OPE potential also admits an $e^{+\sqrt{2}x}$
component which would destroy the renormalizability of
the theory, as it generates divergences which cannot
be absorbed by any finite number of counterterms.
The previous component does not appear however if the ${\rm LO}$ wave functions
have been properly renormalized.
Therefore, what is essential is the power law behaviour of the wave functions,
which dictates the divergence structure of the perturbative integrals
$I^{(\nu)}_{\rho \sigma}$
\begin{eqnarray}
I^{(\nu)}_{\rho\sigma}(k; r_c) &=&
I^{(\nu)}_{0,\rho\sigma}(r_c) + k^2\,I^{(\nu)}_{2,\rho\sigma}(r_c)
\nonumber \\ &+& I^{(\nu)}_{R,\rho\sigma}(k; r_c) \, , 
\end{eqnarray}
with $I^{(\nu)}_{0,\rho\sigma}$ and $I^{(\nu)}_{2,\rho\sigma}$ the divergent
pieces of the integral and $I^{(\nu)}_{R,\rho\sigma}$ the regular piece.
We can regularize the integral $I^{(\nu)}_{\rho \sigma}$ by including
two free parameters
\begin{eqnarray}
\hat{I}^{(\nu)}_{\rho\sigma}(k; r_c) &=&
\lambda^{(\nu)}_{0,\rho\sigma} + \lambda^{(\nu)}_{2,\rho\sigma} k^2 +
I^{(\nu)}_{\rho\sigma}(k; r_c) \, ,
\end{eqnarray}
which are to be fitted to the scattering data of the corresponding phase.
The previous procedure yields a total of six counterterms for regularizing
the ${\rm NLO}$ and ${\rm N^2LO}$ phase shifts.
As in the singlet case, the finite piece of one of these parameters
($\lambda^{(\nu)}_{0,\alpha\alpha}$) is redundant as it only affects
the triplet scattering length $a_t$, which was already fixed at leading order.
In other words, six pieces of data are enough to determine
the ${\rm NLO}$/${\rm N^2LO}$ phase shifts in the triplet.


%

\end{document}